\documentclass[twocolumn,showpacs,preprintnumbers,amsmath,amssymb]{revtex4}

\usepackage{graphicx}
\usepackage{dcolumn}
\usepackage{bm}
\usepackage{textcomp}
\usepackage{wasysym}
\usepackage{stmaryrd}

\def \ee{\end{equation}}
\def \be{\begin{equation}}

\preprint{}

\begin{document}

\title{Quantizing Geometry or Geometrizing the Quantum?}

\author{Benjamin Koch}
 \affiliation{
 Pontificia Universidad Cat\'{o}lica de Chile, \\
Av. Vicu\~{n}a Mackenna 4860, \\
Santiago, Chile \\
}
\date{\today}

\begin{abstract}
The unsatisfactory status of the search for
a consistent and predictive quantization of gravity
is taken as motivation to study the question whether
geometrical laws could be more fundamental
than quantization procedures.
In such an approach the quantum mechanical laws should
emerge from the geometrical theory. A 
toy model that incorporates the idea is presented and 
its necessary formulation in configuration space is emphasized.
\end{abstract}

\maketitle

%
\section{Quantizing Geometry}
The dream of finding a unified description
of all physical phenomena is facing a profound problem:
``{\it{The deep incompatibility between the 
indefinite nature of quantum mechanics
and the rigid geometrical formulation of 
general relativity.}}''
A common assumption is 
that quantum mechanics, as it is usually formulated,
is a fact of ``nature'' and thus it is more fundamental
than general relativity.
Consequently most approaches 
try to apply one of the well defined
quantization procedures to the physical degrees of 
freedom of space-time 
(or to some deeper theory that gives rise to space-time).
Some of the most popular approaches along this line are:
 
{\it{String theory}} is for many the most famous candidate
for a unified theory of nature~\cite{Gross:1985fr}.
It also lead to interesting conjectures about the
relation between certain tree level 
string theories and quantum field theory~\cite{Maldacena:1997re}.
But until today it could not live up
to its promises concerning the uniqueness of what this theory
actually predicts and explains.

{\it{Loop quantum gravity}} is a canonical approach
for the quantization of space-time.
In earlier stages of its development it lead to 
the development of geometrodynamics~\cite{Wheeler:1957mu}
and~\cite{Deser:1976eh} supergravity
 that has very nice
features at the Planck scale~\cite{Rovelli:1997yv}. However,
up to now it was not possible to show that it really
contains general relativity in some 
classical limit~\cite{Ashtekar:2004vs}.

{\it{Causal dynamical triangulation and causal sets}}
are disciplines that earn more and more attention~\cite{Bombelli:1987aa}. They
show the emergence of four dimensional space-time by starting from a discrete
causal structure. Until now those approaches are
limited to asking very basic questions on such as the dimensionality
of space-time but they don't allow to derive 
an effective gravitational action.

{\it{Induced gravity}} theories try to show the emergence of 
curved space-time in a mean field approximation of some underlying
microscopic degrees of freedom~\cite{Sakharov:1967pk}. It is assumed that this
mechanism is similar
to the mechanism that allows to get fluid dynamics from
Bose-Einstein condensation. Up to now those models
manage to mimic some possible features of (quantized) general relativity
but a complete picture is still missing.
An other alternative for emergence of gravity
is based on the idea that the existence of curved space-time
emerges from non-geometric statistical
laws \cite{Jacobson:1995ab,Verlinde:2010hp}.

{\it{Renormalization group}} approaches are
working in the imaginary time formalism.
Given an ultra violet (UV) completion and the existence of a non-trivial fixed
point in the running couplings of the completed
gravitational action this approach might present a
renormalizable version of gravity~\cite{Reuter:1996cp,Litim:2003vp}.
Until now the strict applicability of the imaginary time formalism
and the form of the UV completion are open issues. 

{\it{Anisotropic models}} postulate a different scaling
behavior of space and time in the UV regime,
which allows to construct a power counting renormalizable
theory~\cite{Horava:2009uw} in the UV.
However, recent studies claim that the infrared limit
of the theory is not identical to massless gravity~\cite{Charmousis:2009tc}.

{\it{Further}} research has been done on
asymptotic quantization~\cite{Ashtekar:1981sf},
twistors~\cite{Penrose:1986ca},
non-commutative~\cite{Connes:1996gi,Nicolini:2008aj}
and discretized~\cite{Gambini:2004vz} geometry.

Despite of impressive progress
in some directions, the original
task remains unsolved in all those approaches.

\section{Geometrizing the Quantum}
Given the problems in applying the laws of
quantum mechanics to the geometry of space-time
we want to ask the following question:

``{\it{Could it be that (classical) geometry is more fundamental
than the rules of quantization?}}''
\subsection{Conceptual problems}
Necessarily, answering this question with ``yes''
would mean that the undeniable observable effects of
quantization have to emerge from the deeper
theory (in this case a classical geometrical theory).
Such an approach faces immediately two mayor problems
\begin{itemize}
 \item {\it{Determinism}} \\is,
in contrast to quantum mechanics, part of
most geometric theories (such as general relativity). 
This means for example 
that in causal geometrical theories uncertainties
are just a result of unknown initial conditions, whereas
in standard quantum mechanics they are an irrenunciable
concept.
 \item {\it{Non-locality:}}\\
In principle it is possible to construct deterministic
(hidden variable) theories that are in agreement with the predictions
of quantum mechanics. However, those theories have
to pay a price in order to evade ``no go'' theorems
such as the Bell inequalities~\cite{Bell:1988}. 
They have to contain non-local interactions.
\end{itemize}

\subsection{A conceptual bridge}
There exists a self consistent deterministic
formulation of quantum mechanics, which
also reproduces all typical experimental results
\footnote{Since it can not be distinguished experimentally
from the standard formulation of quantum mechanics
it's probably better to call it an interpretation.}.
It was first suggested by de Broglie, then shown
to be consistent by Bohm~\cite{Bohm:1951,Bohm:1951b}
and later further developed by several authors 
\cite{Holland:1985ud,Bell:1988,Nikolic:2002mi}.
It will be referred to as dBB (de Broglie-Bohm) theory.
In this proposal, the dBB theory will be an essential piece when
building the bridge from classical geometry to
a quantum theory.
We will now shortly present its
formulation for the 
case of a relativistic system of n-bosonic particles
as given by~\cite{Nikolic:2002mi}:
Let $|0\rangle$ be state vector of the 
vacuum and $|n\rangle$ be an arbitrary n-particle state.
The corresponding n-particle wave function is~\cite{Nikolic:2002mi}
\be\label{eq_wavefunction}
\psi(x_1;\;\dots\; ; x_n)= \frac{{\mathcal{P}}_s}{\sqrt{n!}}
\langle 0|\hat{\Phi}(x_1) \dots \hat{\Phi}(x_n)|n\rangle\quad,
\ee
where the $\hat{\Phi}(x_j)$ are scalar Klein-Gordon field operators.
The symbol ${\mathcal{P}}_s$ denotes symmetrization over all 
positions $x_j$ which we will keep in mind but not write explicitly
any more.
For free fields, the wave function (\ref{eq_wavefunction})
satisfies the equation
\be\label{eq_KG0}
\left(\sum_j^n \partial^m_j\partial_{j m}+n \frac{M^2}{\hbar^2} \right)
\psi(x_1;\;\dots\; ; x_n)=0 \quad.
\ee
The mass of a single particle is given by $M$.
The index $j$ indicates on which one of the $n$ particle
coordinates the differential operator has to act
and the index $m$ is the typical space-time index
in four flat dimensions. 
Since quantum mechanics is formulated in terms
of equal time commutation relations, the relevant
equation only contains one single time variable
$\tau=t_1=t_2 \dots t_n$.
Doing this, equation (\ref{eq_KG0}) contains $n$ 
identical time derivatives which can be absorbed
into a single $\partial_t^2$ by redefining 
$t=\tau/\sqrt{n}$. \\
The key step to the dBB interpretation comes from
splitting the wavefunction  up $\psi=P \exp(iS/\hbar)$
and postulating the 
three-momentum of the particle $j$ to be $-\vec{\partial}_{j}S$.
Including this definition one has three coupled real differential equations.
For further convenience the coordinates
for the n particles can be labeled as
\be\label{eq_coordFlat}
x^L=(t,\vec{x}_1,\vec{x}_2,\dots,\vec{x}_n)\quad,
\ee
which also implies the $1+ 3n$ dimensional co- and contravariant
derivatives
$\partial_L$
and $\partial^L$.
Now the three real equations of the dBB theory read
\begin{eqnarray}
\label{eq_KG1L}
2 M Q&\equiv&(\partial^L S)(\partial_{L} S)-n M^2\quad \mbox{with}\\
\nonumber
Q&\equiv&\frac{\hbar^2}{2M} \frac{\partial^L \partial_L P}{P}
\quad,\\
\label{eq_KG2L}
0&\equiv&
\partial_{L} \left(P^2 (\partial^L S) \right)\quad, \\ 
\label{eq_KG3L}
p^L& \equiv& M\frac{d x^L}{ds}\equiv-\partial^L S \quad.
\end{eqnarray}
Applying the total derivative $d/ds\equiv dx^L/ds \partial_L$
to eq.~(\ref{eq_KG3L}) gives a Newtonian type of equation
of motion
\begin{eqnarray}\label{eq_eomL}
\frac{d^2x^L}{ds^2}&=&\frac{
(\partial^N S)(\partial^L\partial_N
S)}{M^2} \quad.
\end{eqnarray}
It is crucial to note that this theory
addresses the two previously mentioned 
conceptual issues and thus makes the dBB 
theory a good
framework for the program of geometrizing the quantum.\\
First, it is deterministic in the sense 
that given initial positions and given initial
field configurations for $S$ and $P$ determine the
final state of the system.\\
Second, it is deeply non-local, because
the functions $S$ and $P$ simultaneously depend
on the positions of all the n-particles.\\
A further remark: the dBB theory contextual and 
therefore not affected by the Kochen-Specker theorem~\cite{Kochen:1976}.
\section{Emergent quantum mechanics}
The idea that quantum mechanics might not be 
fundamental but rather emerge from an underlying classical
system has been proposed in various ways.

\subsection{Various appearances of the idea}
Although the focus of this paper is on
the possible geometric origin of quantum mechanics
it is instructive to give a list of proposals
that point into a similar direction.

{\it{Statistical emergence of quantum mechanics:}}\\
In~\cite{Wetterich:2002fy,Wetterich:2001aj}
it was shown that quantum mechanical correlations
arise when considering finite subsystems of classical
statistical systems with originally infinite degrees of
freedom. An application of this observation
to quantum gravity is perceivable but was not attempted yet.

{\it{Gauge emergence of quantum mechanics:}}\\
Based on a new kind of local gauge transformation
a non-linear field theory has been proposed that
contains quantum field theory as an infrared limit~\cite{Elze:2006fu}.
Also a special classical supersymmetric model was
suggested to give rise to a quantum mechanical system~\cite{Elze:2004by}.
A possible unification with general relativity was not explored yet.

{\it{Dissipative emergence of both, quantum and gravity:}}\\
Dissipative deterministic systems can give rise to
quantum operators and symmetries that are not present
in the original theory at the microscopic
scale~\cite{'tHooft:1999gk,'tHooft:2007xi}.
Further conjecturing that those symmetries
are the ones of diffeomorphism invariance (general relativity)
might give an identikit picture of a future theory of
quantum gravity.

{\it{Geometrical emergence of quantum mechanics:}}\\
The similarity between Weyl geometry and
the structure of quantum mechanical equations 
was first noticed in~\cite{Santamato:1984qe}. 
Other studies in this direction
focused on the Ricci flow~\cite{Carroll:2007zh,Abraham:2008yr}
or on a geometric reduction of the 
dimensionality of space-time~\cite{Gozzi:2006cv,Dolce:2009ce}.
Using local conformal transformations (Weyl geometry) it was
even possible to formulate a geometrical theory that
contains in certain limits both general relativity
and the equations of Bohmian
mechanics~\cite{Shojai:2000us,Bonal:2000zc,Carroll:2004hs}.
The impressive success of those (Weyl geometry) models is limited
to the single particle case because the dBB theory
is only consistent if it also contains the non-local interactions
due to multi particle dynamics.

\subsection{Geometry of configuration space}
It was shown that
existing models for the geometrical emergence
of quantum mechanics are incomplete, since
they can't explain the non-local interactions
in the multi particle dBB theory.
Continuing previous work in this direction~\cite{Koch:2008hn,Koch:2009zz}
a possible way to fill this gap will be presented.

The $1+3 n$ dimensional configuration
space of n-particles with a common time cordinate
will be considered.
Following the notation in eq.~(\ref{eq_coordFlat})
the coordinates in this (possibly curved) space-time will be denoted as
\be\label{eq_coordCurved}
\hat{x}^\Lambda=
(\hat{t},\hat{\vec{x}}_1,\;\dots\;,
\hat{\vec{x} } _n )\quad .
\ee
The toy model for the curvature of this space
will be a single scalar equation which is a $1+3 n$ dimensional analog
to the Nordstrom theory~\cite{Nordstrom:1913a}
\be\label{eq_Nord}
\left.\hat{R}\right|_{\mathcal{S}}=\kappa
\left.\hat{{\mathcal{T}}}_M\right|_{\mathcal{S}}\quad.
\ee
The left hand side contains the Ricci scalar (corresponding
to a metric $\hat{g}_{\Lambda \Sigma}$). The right hand side
contains some coupling constant $\kappa$ and the trace
of the energy momentum tensor $\hat{T}$.
The symbol $|_{\mathcal{S}}$ indicates complete symmetrization
of the terms with respect to the interchange
of two configuration coordinates $\hat{x}_i \leftrightarrow \hat{x}_j,\dots\;$.
Just like in the case of the bosonic Klein-Gordon equation
we will keep this in mind without explicitly writing
it into the following equations.
The symmetrization further fixes the coordinate system for
the four dimensional subspaces and forces
all block diagonal submetrics to be identical
$\hat{g}_i^{\mu \nu}\leftrightarrow \hat{g}_j^{\mu \nu}$.
In order to describe the local conformal part of this theory
separately and for simplification 
one assumes the metric $\hat{g}$ to split up into a
conformal function $\phi(x)$ and a flat part $\eta$
\be\label{eq_gmn}
\hat{g}_{\Lambda \Gamma}=\phi^{\frac{4}{3n-1}} \eta_{L G}\quad.
\ee
The inverse of the metric (\ref{eq_gmn}) is
\be\label{eq_gmnInv}
\hat{g}^{\Lambda \Gamma}=\phi^{-\frac{4}{3n-1}} \eta^{L G}\quad.
\ee
Indices with a lower
Greek and a lower Roman index can be identified
$\hat{\partial}_\Lambda\equiv \partial_L$.
From this follows for example that the adjoint derivatives are not identical,
in both notations
\be\label{eq_derivatives}
\hat{\partial}^\Lambda=\hat{g}^{\Lambda \Sigma}
\hat{\partial}_\Sigma=\phi^{-\frac{4}{3n-1}}
\eta^{L S}\partial_S=
\phi^{-\frac{4}{3n-1}}\partial^L\quad.
\ee

{\it{The geometrical dual to the first dBB equation:}}\\
An Extension of the Hamilton Jacobi stress energy
tensor $\hat{{\mathcal{T}}}_M$ can be
defined 
by subtracting a mass term
$\hat{M}^2$ for every particle
\begin{eqnarray}\label{eq_TMN}
\hat{\mathcal{T}}_M &=& \hat{p}^\Lambda \hat{p}_\Lambda - 
n \hat{M}_G^2\\ \nonumber
&=&(\hat{\partial}^{\Lambda}S_H)(\hat{\partial}_{\Lambda}
S_H)-n \hat{M}_G^2\\ \nonumber
&=&\phi^{\frac{-4}{3n-1}}\left(({\partial}^{L}S_H)({\partial}_{L}
S_H)-n M_G^2\right)\\ \nonumber
&=&\phi^{\frac{-4}{3n-1}}{\mathcal{T}}_M\quad.
\end{eqnarray}
The Hamilton principle function $S_H$ defines the local momentum
$\hat{p}^\Lambda=\hat{M_G}
\,d\hat{x}^\Lambda/d\hat{s}=-\hat{\partial}^\Lambda
S_H$.
Combining (\ref{eq_TMN}), (\ref{eq_derivatives}),  (\ref{eq_gmn}),
and (\ref{eq_Nord}) gives
\be\label{eq_Nord1b}
\frac{12n}{\kappa(1-3n)}\frac{\partial^L\partial_L \phi}{\phi}=
 ({\partial}^{L}S_H)({\partial}_{L}
S_H)-n M_G^2\quad.
\ee
This is exactly the first dBB equation (\ref{eq_KG1L})
if one identifies
\begin{eqnarray}\label{eq_match}
\phi(x)&=&P(x)\quad, \\
S_H(x)&=& S(x)\;\;, \nonumber \\
\kappa&=&\frac{12n}{(1-3n)}/\hbar^2\;\;, \nonumber \\
M^2&=&M_G^2\quad. \nonumber
\end{eqnarray}
Note that the matching conditions demand a negative coupling
$\kappa$.

{\it{The geometrical dual to the second dBB equation:}}\\
In order to find the dual to the second Bohmian
equation one can exploit that the stress-energy tensor (\ref{eq_TMN})
is covariantly conserved
\be
\hat{\nabla}_\Lambda \hat{T}^{\Lambda \Delta}=0\quad.
\ee
This is true if the following relations
are fulfilled
\begin{eqnarray}\label{eq_cons1}
\hat{\nabla}_\Lambda (\hat{\partial}^\Lambda S_H)&=&0 \quad,\\
\label{eq_cons2}
(\hat{\partial}^\Lambda S_H) 
\hat{\nabla}^\Delta (\hat{\partial}_\Lambda S_H)&=&0 \quad, \\
\label{eq_cons3}
(\hat{\partial}^\Lambda S_H) 
\hat{\nabla}_\Lambda (\hat{\partial}^\Delta S_H)&=&0 \quad.
\end{eqnarray}
In addition to the covariant conservation of momentum (\ref{eq_cons1})
and the conservation of squared momentum (\ref{eq_cons2}) the tensor
nature of (\ref{eq_TMN}) also demands (\ref{eq_cons3}).
In order to calculate the covariant derivatives in
(\ref{eq_cons1}-\ref{eq_cons3}),
one needs to know the 
Levi Civita connection
\begin{eqnarray}\label{eq_Levi}
\Gamma^\Sigma_{\Lambda \Delta}&=&\frac{1}{2}g^{\Sigma \Xi}\left(
\partial_\Lambda g_{\Delta \Xi}+\partial_\Delta g_{\Xi \Lambda}-
\partial_\Xi g_{\Lambda \Delta}\right) \\ \nonumber 
&=&\frac{1}{2} \phi^{-\frac{4}{3n-1}}\left[(\partial_L 
\phi^{\frac{4}{3n-1}})\delta^S_D
+(\partial_D  \phi^{\frac{4}{3n-1}})\delta^S_L\right.\\ \nonumber
&&\left.\quad\quad \;\quad-(\partial^S  \phi^{\frac{4}{3n-1}})\eta_{LD}
\right]\quad.
\end{eqnarray}
It is this form of the connection that
gives rise to the non-metricity in Weyl geometry.
Using eq.~(\ref{eq_Levi}), the condition (\ref{eq_cons1}) reads
\be\label{eq_dualKG2L}
\hat{\nabla}_\Lambda (\hat{\partial}^\Lambda S_H)=
\phi^{-\frac{4n}{2n-1}}\partial_L \left[
\phi^2 (\partial^L S_H)
\right]=0\quad.
\ee
With the matching conditions (\ref{eq_match}), 
the above equation is identical to the second
Bohmian equation (\ref{eq_KG2L}).

{\it{The geometrical dual to the third dBB equation:}}\\
According to the Hamilton-Jacobi formalism
the derivatives of the Hamilton principle function ($S_H$)
define the momenta 
\be\label{eq_dualKG3L}
\hat{p}_\Lambda\equiv -(\hat{\partial}_\Lambda S_H)\quad.
\ee
Therefore, with the prescription (\ref{eq_derivatives}) and
the matching condition (\ref{eq_match}) one sees
that the third Bohmian equation (\ref{eq_KG3L})
is fulfilled.

{\it{The geometrical dual to the dBB equation of motion:}}\\
From differential geometry it is known
that the validity of the geodesics equation of
motion results in the conservation of the stress energy
tensor. Nevertheless, it is a good consistency check~\cite{Koch:2009zz}
to explicitly calculate the geodesic equation
\be\label{eq_eom3}
\frac{d^2 \hat{x}^\Lambda}{d\hat{s}^2} +
 \hat{\Gamma}^\Lambda_{\Delta \Sigma}
\frac{d\hat{x}^\Delta}{d\hat{s}}\frac{d\hat{x}^\Sigma}{d\hat{s}}=
\hat{p}^\Lambda \cdot f(\hat{x}) \quad.
\ee
Inserting eq.~(\ref{eq_Levi}) into eq.~(\ref{eq_eom3})
and using the matching conditions eq.~(\ref{eq_match})
the dBB equation of motion (\ref{eq_eomL}) is obtained.
\section{Summary}
It is advocated that ``geometrizing the quantum''
might be a viable alternative to the standard
approaches to quantum gravity.
The main conceptual problems of the new 
approach are discussed.
Using the example of a scalar geometrical toy model 
(incorporating gravity is beyond the scope of this proposal) 
and mapping this model to the dBB interpretation
of the multi particle Klein-Gordon equation, it is shown
how those problems can be evaded.
It is argued that such a mechanism only
can work consistently if the geometrical theory is
formulated in the ($1+3n$ dimensional) configuration space of the system.\\
The author wants to thank J.M. Isidro and D. Dolce for their remarks.



\end{document}